\documentclass[
  journal=pasa,
  manuscript=research-paper,
  year=2022,
  volume=00,
]{cup-journal}

\usepackage{microtype,booktabs}
\usepackage{siunitx}
\usepackage{aas-macros}
\usepackage{amsmath,amssymb}
\usepackage{gensymb} 
\usepackage{hyperref} 
\usepackage{physics}

\hypersetup{
    colorlinks,
    citecolor=blue,
    linkcolor=blue,
    urlcolor=blue,
    pdfauthor={C. P. Lee et al.},
    pdftitle={Spectral analysis of 22 radio pulsars using SKA-Low precursor stations}
}

\defcitealias{Jankowski2018}{JvSK+18}
\newcommand{\dmu}{\,\text{cm}^{-3}\,\text{pc}}
\newcommand{\mJy}{\,\text{mJy}}
\newcommand{\Gauss}{\,\text{G}}
\newcommand{\fig}{Fig. }
\newcommand{\tab}{Table }
\newcommand{\meas}[2]{#1$\,\pm\,$#2} 
\newcommand{\measu}[3]{#1$\,\pm\,$#2$\,$#3} 

\title{Spectral analysis of 22 radio pulsars using SKA-Low precursor stations}

\author{C.~P.~Lee}
\affiliation{Department of Physics and Astronomy, Curtin University, Bentley, WA6102, Australia}
\alsoaffiliation{International Centre for Radio Astronomy Research, Curtin University, Bentley, WA6102, Australia}
\email[C.~P.~Lee]{c.p.lee@student.curtin.edu.au}

\author{N.~D.~R.~Bhat}
\affiliation{International Centre for Radio Astronomy Research, Curtin University, Bentley, WA6102, Australia}

\author{M.~Sokolowski} 
\affiliation{International Centre for Radio Astronomy Research, Curtin University, Bentley, WA6102, Australia}

\author{N.~A.~Swainston}
\affiliation{International Centre for Radio Astronomy Research, Curtin University, Bentley, WA6102, Australia}

\author{D.~Ung}
\affiliation{International Centre for Radio Astronomy Research, Curtin University, Bentley, WA6102, Australia}

\author{A.~Magro} 
\affiliation{Institute of Space Sciences and Astronomy, University of Malta, Msida, Malta}

\author{R.~Chiello}
\affiliation{University of Oxford, Denys Wilkinson Building, Oxford, UK}

\doi{XX.XXXX/pasa.XXXX.XX}

\received {dd Mmm YYYY}
\revised  {dd Mmm YYYY}
\accepted {dd Mmm YYYY}
\published{dd Mmm YYYY}

\keywords{instrumentation: interferometers -- methods: observational -- pulsars: general -- stars: neutron}

\begin{document}

\begin{abstract}
We present the first observational study of pulsars performed with the second-generation precursor stations to the low-frequency component of the Square Kilometre Array (SKA-Low): the Aperture Array Verification System 2 (AAVS2) and the Engineering Development Array 2 (EDA2). Using the SKA-Low stations, we have observed 100 southern-sky pulsars between 70--350$\,$MHz, including follow-up observations at multiple frequencies for a selected sample of bright pulsars. These observations have yielded detections of 22 pulsars, including the lowest-frequency detections ever published for 6 pulsars, despite the modest sensitivity of initial system where the recording bandwidth is limited to $\sim\SI{1}{\MHz}$. By comparing simultaneous flux density measurements obtained with the SKA-Low stations and performing rigorous electromagnetic simulations, we verify the accuracy of the SKA-Low sensitivity simulation code presented in \citet{Sokolowski2022}. Furthermore, we perform model fits to the radio spectra of the detected pulsars using the method developed by \citet{Jankowski2018}, including 9 pulsars which were not fitted in the original work. We robustly classify the spectra into 5 morphological classes and find that all but one pulsar exhibit deviations from simple power-law behaviour. These findings suggest that pulsars with well-determined spectra are more likely to show spectral flattening or turn-over than average. Our work demonstrates how SKA-Low stations can be meaningfully used for scientifically useful measurements and analysis of pulsar radio spectra, which are important inputs for informing pulsar surveys and science planned with the SKA-Low.
\end{abstract}

\section{INTRODUCTION}
\label{sec:int}
The radio spectra of pulsars offer unique insights into the nature of the mysterious pulsar emission mechanism, and for this reason have been the subject of extensive study for many decades \citep[e.g.][]{Sieber1973,Malofeev1980,Izvekova1981,Lorimer1995,Maron2000}. Furthermore, studies of pulsar spectra are useful resources for planning surveys of the Galactic pulsar population with the Square Kilometre Array (SKA). In particular, spectral modelling can be used to inform pulsar population studies, which play a key role in estimating the yields of future surveys \citep[e.g.][]{Xue2017,Keane2015}. The SKA is expected to discover thousands of new pulsars and millisecond pulsars (MSPs), which will be critically important in developing pulsar timing arrays (PTAs) for the detection of ultra-low-frequency gravitational waves \citep[e.g.][]{Manchester2013}. PTAs are one of a broad range of high-profile applications that have led to pulsars being ranked as a headline science for the SKA \citep[e.g.][]{Janssen2015}.

It is well known that the majority of pulsars exhibit steep power-law spectra ($S_\nu\propto\nu^{\alpha}$; where $\alpha$ is the spectral index and $S_\nu$ is the flux density at frequency $\nu$) with a mean spectral index of $\expval{\alpha}=\num{-1.60\pm 0.03}$ \citep{Jankowski2018}. However, the radio spectra of some pulsars have been observed to reach a peak in flux density at low radio frequencies (a spectral `turn-over'), which is most commonly attributed to either synchrotron self-absorption \citep{Sieber1973} or thermal free-free absorption \citep{Malov1979}. Interestingly, studies of the MSP population have found that they do not turn over like long-period pulsars \citep[e.g.][]{Kramer1999,Kuzmin2001,Toscano1998}, but rather continue as power-laws to the lowest observable frequencies \citep[e.g.][]{Erickson1985}. Only a handful of MSPs have shown hints of turning over \citep[e.g.][]{Kuniyoshi2015,Dowell2013}, all of which are predicted to peak well below \SI{100}{\MHz}, which suggests that turn-overs in MSPs may occur at much lower frequencies than long-period pulsars. Deviations from a simple power-law (such as spectral flattening and low-frequency turn-over) are referred to as spectral features. \citet[][hereafter \citetalias{Jankowski2018}]{Jankowski2018} performed a study of the spectral properties of 441 pulsars, and found that 21\% of pulsars show spectral features, increasing to 44\% for pulsars with good low-frequency coverage (i.e. more than two data points below \SI{600}{\MHz}). Building upon this work, \citet{Johnston2021} classified an additional 44 pulsars, 20\% of which show spectral features.

Unfortunately, the vast majority of catalogued pulsars lack reliable flux density measurements below \SI{400}{\MHz}, which puts poor constraints on spectral models at these frequencies. A recent version of the ATNF pulsar catalogue\footnote{\url{https://www.atnf.csiro.au/people/pulsar/psrcat/}} \citep[version 1.67, released in March 2022;][]{PSRCAT} shows that $\sim70\%$ of the over 3300 known pulsars do not have flux density measurements available below \SI{400}{\MHz}. This is despite the emergence of the latest generation of low-frequency aperture array telescopes such as the Murchison Widefield Array \citep[MWA;][]{Tingay2013,Wayth2018}, the LOw-Frequency ARray \citep[LOFAR;][]{vanHaarlem2013,Stappers2011}, and the Long Wavelength Array \citep[LWA;][]{Taylor2012,Ellingson2013}, which have led to the publication of flux densities for a considerable number of known pulsars at low frequencies \citep[e.g. ][]{Murphy2017,Xue2017,Bon2020,Bilous2020,Bilous2016,Stovall2015}.

In general, pulsar flux density measurements can be obtained in two ways: (1) measurements of pulsed emission from beamformed detections, which often have inaccurate absolute flux calibrations and therefore limit constraints on spectral models; and (2) measurements from interferometric continuum images, which are less sensitive, but provided that data are recorded for a large field-of-view \citep[e.g. the MWA Voltage Capture System; VCS;][]{Tremblay2015}, can be robustly calibrated by virtue of the ability to perform in-field calibration against hundreds of other calibrator sources. Continuum measurements may also suffer from source confusion (i.e. blending with other sources in the field), depending on the spatial resolution. The vast majority of flux density measurements are made from pulsed emission, however the advent of widefield low-frequency aperture arrays has enabled continuum flux density measurements to be made for a small number of pulsars \citep[e.g.][]{Murphy2017}.

The low frequency component of the SKA (SKA-Low), to be built at the Murchison Radio-astronomy Observatory (MRO) in Western Australia, will operate between 50--\SI{350}{\MHz}, and will be capable of providing substantially improved constraints on pulsar emission properties at low frequencies, including the nature of the spectra. Currently operational at the MRO are two second-generation prototype SKA-Low stations: the Aperture Array Verification System 2 \citep[AAVS2;][]{AAVS2confpaper,AAVS2} and the Engineering Development Array 2 \citep[EDA2;][]{EDA2}. Aside from providing early insights into the capabilities of SKA-Low technologies, these stations are being used to prepare for science to be conducted with the SKA-Low. The stations are particularly useful tools for performing pulsar monitoring due to the real-time beamforming capability, which yields more manageable data volumes and enables faster data processing turn-around times in comparison to the MWA-VCS.

In this work, we demonstrate the early pulsar detection capabilities of the AAVS2 and EDA2 by performing low-frequency observations of a selection of bright southern-sky pulsars, measuring flux densities for the detected pulsars, and modelling their radio spectra. In Section \S\ref{sec:obs}, we describe the facilities, target selection, and observations. In Section \S\ref{sec:dat}, we outline the flux density measurement and calibration methodology, as well as the methods used to model and classify the pulsar spectra in a robust and unbiased way. In Section \S\ref{sec:res}, we present the results of a shallow all-sky census and multi-frequency follow-up observations, and report flux density measurements and spectral classifications for the detected pulsars. Finally, we summarise and give conclusions in Section \S\ref{sec:con}.

\begin{figure*}
    \centering
    \includegraphics[width=\linewidth]{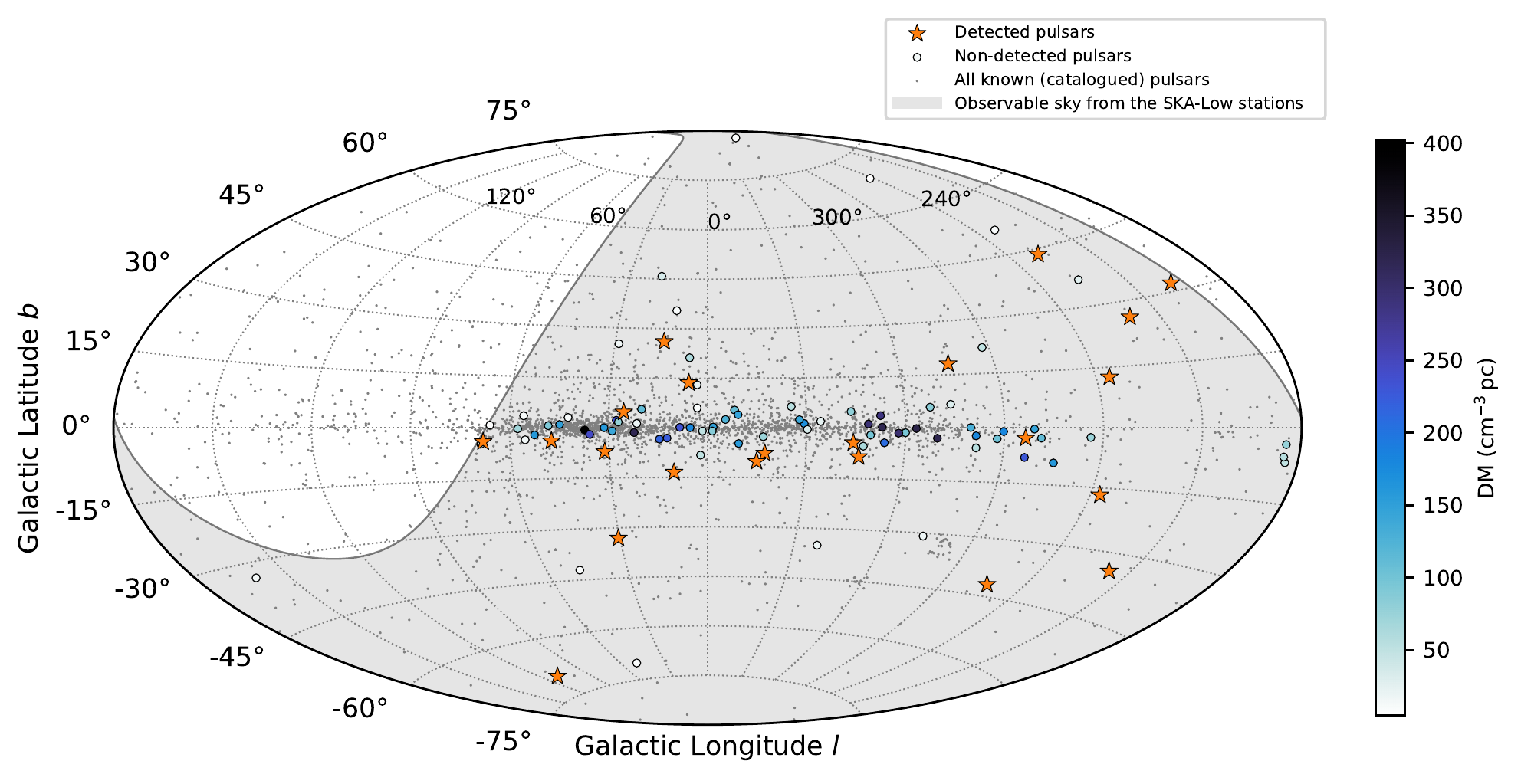}
    \caption{Aitoff projection of the sky showing the distribution of all known pulsars from the ATNF pulsar catalogue (grey dots), the confirmed pulsar detections using the initial capabilities of the SKA-Low stations (orange stars), and the pulsars observed but not detected (colour filled circles) in Galactic coordinates. The colour scale indicates the DM of the non-detected pulsars. The grey shaded region is the observable sky with the SKA-Low stations (declination $\delta<+\SI{30}{\degree}$).}
    \label{fig:skycover}
\end{figure*}

\section{OBSERVATIONS}
\label{sec:obs}
\subsection{SKA-Low prototype stations}
\label{subsec:fas}
\subsubsection{The AAVS2}
The AAVS2 is a low-frequency aperture array that comprises 256 log-periodic dual-polarisation SKALA-4.1 antennas distributed pseudo-randomly over a circular ground plane with a station diameter of $\sim\SI{38}{\m}$. A description of the AAVS2 is provided in \citet{AAVS2confpaper}, and a characterisation of its performance can be found in \citet{AAVS2}. The antenna arrangement replicates the EDA2 and the previous generation stations---the AAVS1~\citep{2021A&A...655A...5B} and the  EDA1~\citep{2017PASA...34...34W}---to enable direct comparisons to be made, however the diameter of the AAVS2 was upscaled to accommodate the larger antennas. The 50--\SI{350}{\MHz} observing band is split into coarse channels which are separated by $\approx\SI{0.781}{\MHz}$ and oversampled up to a bandwidth of $\approx\SI{0.926}{\MHz}$. For pulsar observations, the station forms a phased-array beam by coherently summing the complex voltages of each array element. The station beam can be electronically steered to an arbitrary pointing direction by applying phase delays to each of the antenna signals using digital beamforming implemented in the firmware of Tile Processing Units \citep[TPMs;][]{Naldi2017}. The beam response varies as a function of both frequency and pointing direction, with the sensitivity generally being highest at the zenith. The initial system, which was employed for the work presented in the paper, allowed data capturing from a single coarse channel at a maximum time resolution of $\approx\SI{1.08}{\us}$. Since then, the system has been upgraded for a much larger bandwidth ($>32$ coarse channels), the testing and commissioning of which is in progress.

\subsubsection{The EDA2}
Developed as a comparator to the AAVS2, the EDA2 uses the same analogue and digital signal chain. However, the EDA2 makes use of MWA-style dual-polarisation bowtie dipoles with modified low-noise amplifiers (LNAs) which extend the sensitivity down to $\sim\SI{50}{\MHz}$ to replicate the frequency range of the AAVS2. Details of the design, calibration, and performance of the EDA2 are given in \citet{EDA2}. Its predecessor, the EDA1, was used in \citet{Bhat2018} to observe two MSPs at low frequencies between $\sim50$--\SI{300}{\MHz}, giving an early demonstration of pulsar science with an SKA-Low prototype station.

\subsection{Beamforming: SKA-Low stations vs. the MWA}
The MWA is a low-frequency aperture array telescope which serves as a precursor to the SKA-Low and has been operational at the MRO since 2013 \citep{Tingay2013}. The system is capable of producing high-time- and high-frequency-resolution observations, which has enabled a wide range of meaningful low-frequency pulsar science to be performed, from studies of emission properties, multipath propagation, and pulsar timing, to the first pulsar discoveries with the MWA \citep[e.g.][]{McSweeney2017,Meyers2018,Bhat2018,Kaur2019,Kirsten2019,Swainston2021,McSweeney2022}. The SKA-Low stations have $\sim20\%$ of the MWA's effective collecting area (as indicated by simulations at \SI{160}{\MHz}), and $\sim3\%$ of the bandwidth, giving them $<4\%$ of the sensitivity. However, the MWA performs offline incoherent and tied-array beamforming on tile voltages recorded at 28$\,$TB$\,$h\textsuperscript{-1} using the high-time-resolution VCS backend \citep[][]{Tremblay2015}. In comparison, the SKA-Low stations record beamformed data at only 12$\,$GB$\,$h\textsuperscript{-1} (for a single coarse-channel). This makes the data easier to process due to the beamforming having already been performed in real-time, effectively enabling the stations to operate like single-dish telescopes. The more manageable data volumes and faster data processing turn-around times in comparison to the MWA-VCS make the stations useful tools for performing data-intensive science such as high-cadence pulsar monitoring.

\subsection{Selection of target pulsars}
\label{subsec:tar}
The AAVS2 and EDA2 are relatively new instruments that can be employed for conducting a shallow all-sky census (i.e. sensitive primarily to bright pulsars, due to the limited sensitivity of the initial system). For this work, we compiled a modest sample of bright southern-sky pulsars with version 1.64 of the ATNF pulsar catalogue using the following constraints: (1) a declination limit of $\delta<+30\degree$ (J2000), the same as the MWA; (2) a mean flux density at \SI{400}{\MHz} of $S_{400}>40\mJy$ or a mean flux density at \SI{1400}{\MHz} of $S_{1400}>5\mJy$; and (3) a dispersion measure (DM) cut-off of $450\dmu$. Since the vast majority of catalogued pulsars do not have flux densities reported below \SI{400}{\MHz}, the flux density constraints were calculated by extrapolating higher-frequency flux densities down to \SI{150}{\MHz} using a nominal spectral index of $\alpha=-1.6$. Based on the expected sensitivities of the SKA-Low stations, we chose $200\mJy$ to be the minimum predicted flux density at \SI{150}{\MHz}. Finally, the DM cut-off was chosen to eliminate pulsars with very high DMs and to limit the number of pulsars to 100. The pulsars observed in this work are shown on a Galactic sky map in \fig\ref{fig:skycover}.

The SKA-Low stations are capable of observing nearly three octaves of frequency range, which makes them useful instruments for studying pulsar emission properties over a large and uncommonly observed part of the radio spectrum. To this end, nine pulsars were selected as targets for follow-up observations at multiple frequencies. To select these targets, the successful detections from the initial all-sky census were ranked in order of signal-to-noise (S/N) ratio to give preference to the brightest pulsars, and daytime transiting pulsars were excluded. One of the selected pulsars (J1820$-$0427) did not yield any further detections from follow-up observations.

\subsection{Observations}
\label{subsec:obs}
The pulsars selected for follow-up observations were observed at 18 different centre frequencies from 70.3--351.6$\,$MHz. The observations for each pulsar were conducted over the course of a single night, one at a time, starting at the lowest frequency channel and incrementing to the next highest selected channel with each consecutive observation. The follow-up observations were 15$\,$min in duration (half that of the 30$\,$min census observations), separated by 2$\,$min intervals\footnote{Since the system is in the early stages of development, a conservative window was given to allow for communications overhead between observations.}. PSRs J0835$-$4510, J0953+0755, and J1645$-$0317 were simultaneously observed with the EDA2 and the AAVS2, and the remaining pulsars were observed solely with the EDA2 while the AAVS2 was taken offline for a hardware upgrade. Further observations were taken of PSR J0835$-$4510 to assess the accuracy of the beam sensitivity simulations across a range of elevations. Data collected with the stations were saved onto dedicated on-site servers and later transferred to the Pawsey Supercomputing Centre\footnote{\url{https://pawsey.org.au/}} where they were stored and processed.

\begin{figure*}
    \centering
    \includegraphics[width=\linewidth]{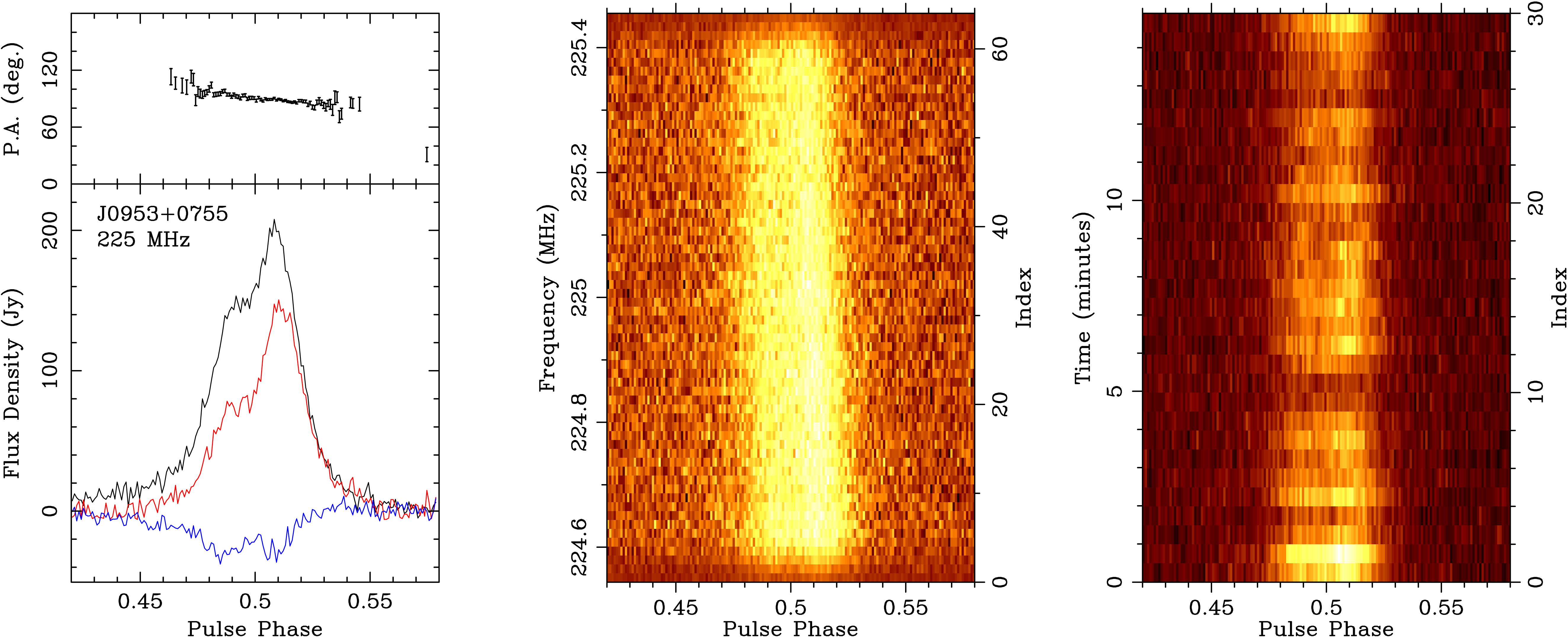}
    \caption{A bright detection of PSR J0953+0755 in 15$\,$min of data ($\sim\text{3500}$ pulses) collected with the EDA2. Plots are shown at a time resolution of $\sim\SI{250}{\us}$. \textit{Left:} Average pulse profile showing polarisation components. The black line is the total intensity (Stokes $I$), the red line is the linearly polarised intensity (Stokes $\sqrt{Q^2+U^2}$), and the blue line is the circularly polarised intensity (Stokes $V$). The polarisation position angle curve is plotted in the top panel. \textit{Centre:} Flux density plotted as a function of frequency and pulse phase. \textit{Right:} Flux density plotted as a function of time and pulse phase with \SI{30}{\s} time integrations.}
    \label{fig:J0953_profile}
\end{figure*}

Despite the limited recording bandwidths ($\sim\SI{1}{\MHz}$), the SKA-Low stations have demonstrated the capabilities to make high-S/N detections of several bright pulsars. \fig\ref{fig:J0953_profile} shows an EDA2 observation of PSR J0953+0755: a short-period ($P\approx\SI{0.25}{\s}$), low-DM ($\approx 2.96\dmu$) pulsar which is known to exhibit strong variability in its observed intensity due to interstellar scintillation \citep{Bell2016,2021PASA...38...23S}. The observation was taken while the pulsar was fortuitously in a brightened state, which resulted in a peak flux density of over 200$\,$Jy. Single pulses have been detected in this observation, and their further investigation will be left for a future publication. The frequency versus phase plot shows negligible dispersion sweep ($\Delta t\approx\SI{2}{\ms}$) due to the small observing bandwidth and low DM, but it can still be discerned in the figure.

Pulsars are very highly polarised radio sources, and polarimetric analysis of their emission can provide useful information about their beam geometry and the intervening interstellar medium (ISM). In principle, the SKA-Low stations are capable of producing full-Stokes pulse profiles which can be used to assess the polarimetric calibration of the stations \citep[e.g.][]{Xue2019}. The pulse profile in \fig\ref{fig:J0953_profile} shows the linear and circular polarisations and the corresponding linear polarisation position angle curve of the bright observation of PSR J0953+0755. Although the observation has not been derotated, this pulsar has a small rotation measure ($\text{RM}\approx\SI{-0.66}{\radian\per\m\squared}$), and over the small recording bandwidth the effect of Faraday rotation is minimal. \cite{McSweeney2020} published a polarisation pulse profile of this pulsar using MWA data at \SI{152}{\MHz} which shows first-order agreement with the EDA2 observation.

\section{DATA PROCESSING AND ANALYSIS}
\label{sec:dat}

\subsection{Integrated pulse profiles}
\label{subsec:prof}
Raw voltages were saved into binary files containing 5$\,$min of data each, which were subsequently merged and prepended with a header containing the pulsar ephemeris, period, DM, and other observing metadata. The data were then processed into archive files using the {\sc dspsr} software package\footnote{\url{http://dspsr.sourceforge.net}} \citep{DSPSR}. The observations were coherently dedispersed, and a 256-point convolving filterbank was run to fine-channelise the coarse channel. The time series was folded into 64 phase bins for initial detections and 256 phase bins for flux measurements to increase the number of off-pulse bins for calibration. The {\sc psrchive} data analysis library\footnote{\url{http://psrchive.sourceforge.net}} \citep{PSRCHIVE} was used to average the data in time and frequency and construct the pulse profiles shown in \fig\ref{fig:profiles}.

\subsection{Calculation of mean flux densities}
\label{subsec:calcflux}
To calculate a mean flux density $S$, the $N_\text{on}$ on-pulse phase bins were added together and the result was normalised by the pulsar period $P$, i.e.
\begin{equation}
    S = \frac{1}{P}\sum_{i=1}^{N_\text{on}} \Lambda I_i \Delta t,
\end{equation}
where $I_i$ is the Stokes $I$ value of the $i$-th on-pulse phase bin, $\Delta t$ is the bin width, and $\Lambda$ is a calibration constant. The on-pulse was estimated using the formula
\begin{equation}
    n_\text{r,l}=\qty(n_\text{max}\pm k N_{3\sigma})\mod N,
\end{equation}
where $n_\text{r,l}$ are the bin indexes of the right and left on-pulse boundaries, $n_\text{max}$ is the index of the bin with the maximum Stokes $I$ value across the profile, $N_{3\sigma}$ is the number of phase bins with a Stokes $I$ value more than three standard deviations from the mean, $k$ is a scaling constant, and $N$ is the total number of phase bins. The $N_{3\sigma}$ value gives an estimate for the number of on-pulse phase bins, which is most accurate for profiles with short duty cycles, but typically underestimates the on-pulse for profiles with larger duty cycles. The value of $k$ (typically 2 or 3) was selected to ensure that all on-pulse estimates were generous enough to pick up any residual emission. Each on-pulse estimate was checked visually to ensure sensibility.

The pulse profile of PSR J0835$-$4510 is heavily scattered at low frequencies, with the scattering tail dominating the off-pulse region at frequencies below around \SI{300}{\MHz}. This makes it difficult to calibrate reliably using measurements of the off-pulse noise. For the lowest-frequency detections, we processed the average pulse profile with a DM of zero to determine an upper-limit of the background noise level. We then used this estimate to calibrate the profile and calculate the flux density. Since these measurements are likely underestimates, they were excluded from the spectral fits. This method works under the condition that the dispersion sweep exceeds the pulse period. For a single coarse-channel, this condition is only met for J0835$-$4510 below $\SI{185}{\MHz}$.

\subsection{Calibration procedure}
\label{subsec:cal}
Calibration of the flux densities required estimations of the system-equivalent flux density (SEFD; i.e. a measure of the system sensitivity) in the pointing direction and at the frequency of each observation. The SEFD is defined as
\begin{equation}
    \text{SEFD} = \frac{T_\text{sys}}{G},
\end{equation}
where $T_\text{sys}$ is the system temperature and $G$ is the gain. At low frequencies, $T_\text{sys}$ is dominated by the sky background, which is a strong function of frequency \citep[$\propto\nu^{-2.55}$;][]{Lawson1987}. The gain is also a function of frequency due to its dependence on the effective collecting area of the array elements. To calculate the SEFD, the beam response was simulated using {\sc python} code\footnote{\url{https://github.com/marcinsokolowski/station_beam}} detailed in \citet{Sokolowski2022}. The simulation code finds $T_\text{sky}$ and $G$, calculates the SEFD in the $X$ and $Y$ polarisations, and combines them in quadrature according to
\begin{equation}\label{eq:sefd}
    \text{SEFD}_I = \frac{1}{2}\sqrt{\text{SEFD}_{XX}^2+\text{SEFD}_{YY}^2},
\end{equation}
where $\text{SEFD}_I$ is the SEFD in the Stokes $I$ polarisation. Equation \eqref{eq:sefd} assumes an unpolarised target source, and is only strictly valid in the cardinal planes ($\phi=0\degree,90\degree$; where $\phi$ is the azimuth angle), which leads to errors at low elevations \citep{Sutinjo2021}. The radiometer equation was used to calculate the standard deviation of the background noise expected in images of the sky,
\begin{equation}
    \sigma_\text{sim}=\frac{\text{SEFD}_I}{\sqrt{n_pt_\text{int}\Delta\nu}},
\end{equation}
where $n_p=2$ is the number of polarisations, $t_\text{int}$ is the integration time, and $\Delta\nu$ is the bandwidth. We have assumed $\sigma_\text{sim}$ to be equal to the off-pulse noise of beamformed observations. By comparing $\sigma_\text{sim}$ to the standard deviation of the off-pulse noise in the uncalibrated pulse profile ($\sigma_\text{off}$), a flux density calibration constant $\Lambda$ was calculated:
\begin{equation}
    \Lambda = \frac{\sigma_\text{sim}}{\sigma_\text{off}}.
\end{equation}

The results from the sensitivity simulation code were compared against a full-electromagnetic simulation of the EDA2 station beam in the commercial simulation package FEKO\footnote{\url{https://www.altair.com/feko}}. It was verified that the simulation code consistently calculated SEFDs within $\sim30\%$ of the FEKO simulation. Due to limitations in time and resources, the flux densities could not all be calibrated against FEKO simulations. Instead, the relative error in the SEFD estimates from the simulation code was accounted for in the uncertainty calculations.

A set of observations were performed at a range of elevations from $\sim20-70\degree$ for a bright pulsar (J0835$-$4510) to test the flux calibration far from the zenith. The error was negligible at elevations greater than $\sim40\degree$, so no elevation-dependent corrections were implemented.

\begin{figure*}[!ht]
    \centering
    \includegraphics[width=\linewidth]{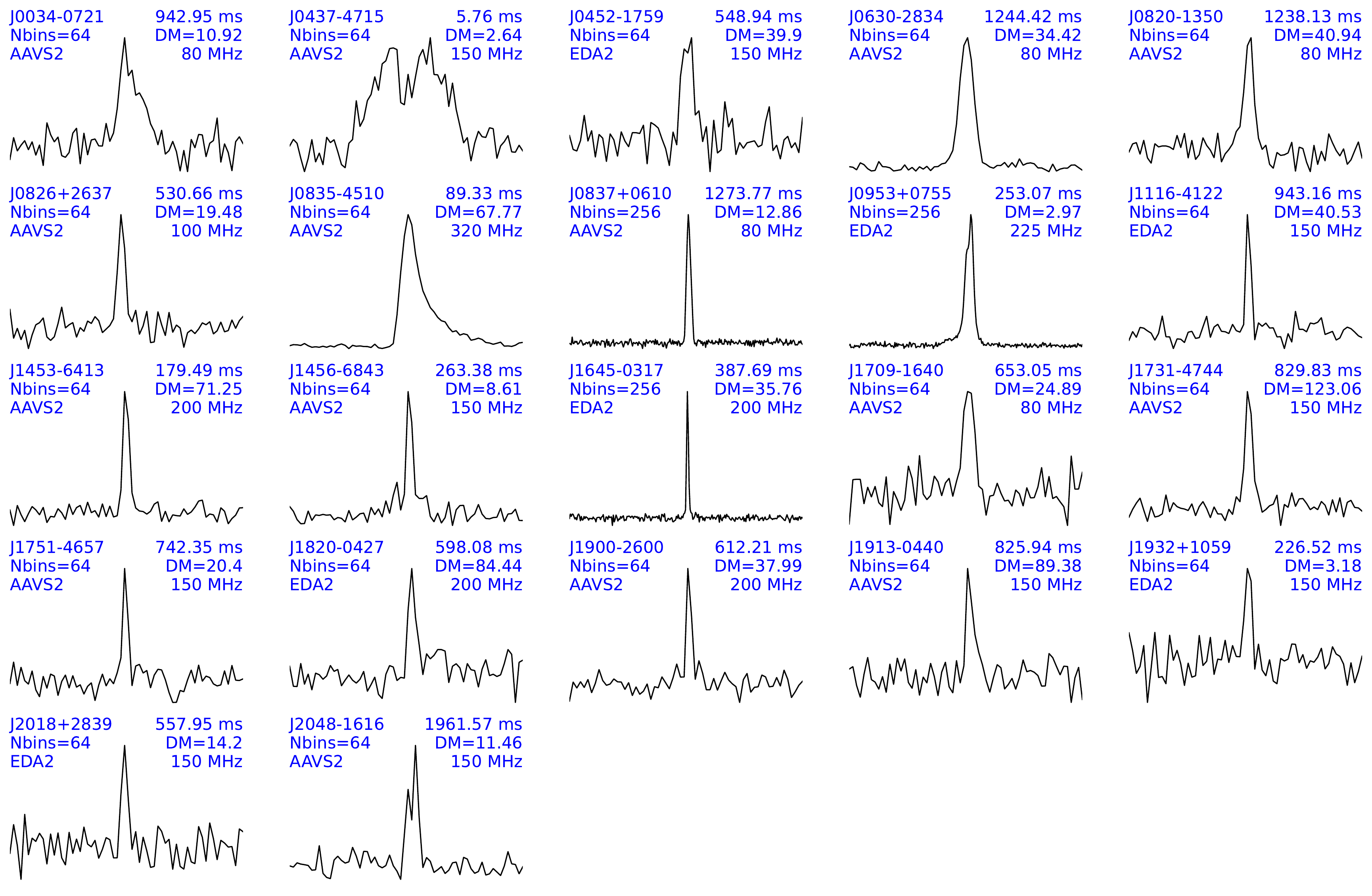}
    \caption{Integrated pulse profiles for the 22 detected pulsars with the SKA-Low stations, including one MSP (J0437$-$4715). The period, number of phase bins, DM (in units of $\!\dmu$), telescope, and observing frequency are also indicated in each panel. In the case of pulsars with multiple detections available, we have shown the best detection. Of the displayed profiles, 15 are AAVS2 detections and 7 are EDA2 detections. In most cases, profiles are shown with 64 phase bins, however high-S/N profiles (e.g. J0953+0755) are shown at a higher time resolution with 256 phase bins.}
    \label{fig:profiles}
\end{figure*}

\subsection{Compiling spectral data}
Flux density measurements are often reported in pulsar discovery papers and census papers, however compiling these resources can often prove time consuming. In the interest of streamlining this process for future work, we have developed an open-source database of published spectral data called {\sc pulsar spectra}\footnote{\url{https://github.com/NickSwainston/pulsar_spectra}}, which will be described in full in a future publication \citep{Swainston2022}. A complete list of publications included in the database can be found in the documentation\footnote{\url{https://pulsar-spectra.readthedocs.io/en/latest/catalogue.html##papers-included-in-our-catalogue}}. The current version of the database (v1.4) comprises the majority of pulsar flux density measurements available at radio frequencies, with particular emphasis on compiling low-frequency measurements, where spectra often exhibit flattening or curvature. For the pulsars analysed in this work, our database is nearly up-to-date.

\subsection{Modelling the pulsar spectra}
\label{subsec:mod}
Robust modelling of data combined from the literature is difficult due to the measurements being made with different telescopes and calibration procedures that have varying levels of reliability. In many cases, this introduces mismatches in systematic errors, which can make modelling the data complicated. Our parameter fitting method is based on the work of \citetalias{Jankowski2018}, who addressed these issues using methods from robust regression and information theory. The spectral modelling code used in this work is included in version 1.4 of {\sc pulsar spectra}. Further details regarding the modelling method can be found in \citetalias{Jankowski2018} and the references within, and a full description of our implementation of the method will be given in Swainston et al. (submitted). A high-level summary of the modelling method is given below.

We used a Gaussian likelihood function, which quantifies the probability of a dataset given a set of model parameters. The likelihood was modified using the Huber loss function \citep{Huber1964}, which penalises outlier data points by reducing their contribution to the model fit. A robust cost function was then derived by taking the negative logarithm of the likelihood function. The optimal parameters were found by minimising the cost function, which is equivalent to maximising the likelihood function.

The cost function was minimised using the {\sc migrad} algorithm from the {\sc minuit2} minimisation library, which was integrated into {\sc pulsar spectra} via the {\sc python} bindings in {\sc iminuit}\footnote{\url{https://github.com/iminuit/iminuit}} \citep{iminuit}. {\sc migrad} converges towards a local minimum using a combination of Newton steps and gradient descents, and is called a set number of times before returning the best-fitting model parameters \citep{James1975}. The quality of the minimisation was verified by ensuring that the best-fitting parameters satisfied a convergence criterion, defined in terms of a specified tolerance value. The uncertainties were computed using {\sc hesse}, an error calculator which computes the covariance matrix for the best-fitting parameters and determines the 1$\sigma$ uncertainties as the square-root of the diagonal elements.

The five analytical models identified in \citetalias{Jankowski2018} were used as a representative sample of the various morphologies reported in the literature. These models are simple power-law, broken power-law, log-parabolic spectrum, power-law with a low-frequency turn-over, and power-law with a high-frequency hard cut-off. Following \citetalias{Jankowski2018}, the parameter defining the smoothness of the low-frequency turn-over was restricted to a value of $0<\beta\leq 2.1$. In each spectral model, the frequency was scaled by a reference frequency which was calculated as the geometric mean of the minimum and maximum frequencies in the spectrum.

In contrast with \citetalias{Jankowski2018}, we allowed the spectral index of the high-frequency cut-off model to be a fit parameter. This modification to the application of the model is in accordance with the original work \citep{Kontorovich2013}, and enables the model to be fitted to a larger sample of spectra.

The best-fitting model was determined using the Akaike information criterion (AIC), which measures how much information the model retains about the data without overfitting. The model which resulted in the lowest AIC was selected as the best-fitting model. Furthermore, the probability that the selected best-fitting model is the true best-fitting model amongst those tested, $p_\text{best}$, was calculated as the inverse sum of the likelihoods of each model relative to the best-fitting model.

\section{RESULTS AND DISCUSSION}
\label{sec:res}
\subsection{Pulsar detections}
\label{subsec:detections}
Detections were made of 22 pulsars using data collected with the SKA-Low stations. The non-detection of the remaining 78 pulsars is due to a range of factors, including overestimated flux densities and interstellar scattering. Our initial estimates for the flux densities at \SI{150}{\MHz} (i.e. $S_{150}$) were made using a rudimentary simple power-law assumption (see \S\ref{subsec:tar}) with single flux density measurements from the ATNF pulsar catalogue (i.e. $S_{400}$ or $S_{1400}$). In light of the development of {\sc pulsar spectra}, we performed spectral fits for all of the non-detected pulsars to estimate how many of the non-detections can be attributed to overestimated flux densities due to spectral flattening or turn-over. Only 29 ($\sim 40\%$) of non-detected pulsars are above the sensitivity limit of the stations (i.e. $S_{150}>200\mJy$, conservatively). Out of these, 14 have high DMs ($\geq100\dmu$). Although DM smearing cannot be a factor (the observations were coherently de-dispersed), high-DM pulsars are much more likely to experience scattering by the ISM \citep[e.g.][]{Bhat2004}, which can reduce the S/N and hence the detection rate of these pulsars. The remaining 15 bright ($S_{150}>200\mJy$) low-DM ($<100\dmu$) non-detected pulsars include the Crab pulsar (PSR J0534+2200), which was likely not detected due to being heavily scattered at low-frequencies \citep[e.g.][]{Meyers2017}; five pulsars whose $S_{150}$ values were likely overestimated due to lack of spectral data below $\sim\SI{600}{\MHz}$ to reveal the presence of any spectral flattening or turn-over; and nine that cannot presently be accounted for.

Detections were made across the frequency range 70.3--\SI{351.6}{\MHz} with both stations, although most pulsars were only detected near the middle of the frequency band where the sensitivity is highest. The bright pulsars J0953+0755 and J1645$-$0317 were detected across the majority of the frequency band, with only a couple of non-detections near the low and high ends of the frequency range. Radio-frequency interference (RFI) affected a small number of observations, which were either excised or discarded depending on the extent of the contamination. Five pulsars were detected by only one of the stations. The non-detections were eventually traced to either a malfunctioning data capturing system or RFI. Therefore, 17 pulsars were simultaneously detected by both stations and direct comparisons can be made between these detections.

The highest-DM pulsar in the detected sample was PSR J1731$-$4744 ($\text{DM}\approx\SI{123.06}{\per\cm\cubed}\text{pc}$), which was detected down to \SI{132.8}{\MHz}. It is also particularly noteworthy that PSR J0437$-$4715 was detected by both stations---this MSP's exquisite timing stability has placed it as a high-profile target for pulsar timing arrays \citep[e.g.][]{Manchester2013}, and its high flux density enables detailed studies into the remarkable evolution of its integrated pulse profile \citep{Bhat2018}.

A gallery of average pulse profiles showing the best detection of each pulsar is displayed in \fig\ref{fig:profiles}. The stations have demonstrated the capability to make relatively high-S/N detections for several bright pulsars. Particularly notable are the detections of PSR J0953+0755, the strongest of which exceeded a S/N of 250 (with 128 phase bins). Pulse profiles with a high S/N were re-processed at a higher time resolution.

The S/N ratio measured by each station during a simultaneous observation can be used to directly compare the station performances. For each observation, the S/N ratio was estimated using {\sc psrchive}. On average, the AAVS2 produced higher-S/N detections than the EDA2. This was most apparent near the upper end of the observing band where the log-parabolic SKALA-4.1 antennas used by the AAVS2 are noticeably more sensitive than the MWA-style bowtie dipoles used by the EDA2.

Whilst the low sensitivity of the stations is often insufficient to study the features of pulse profiles, the detections of PSRs J0437$-$4715 and J2048$-$1616 each show two distinct pulse components that are consistent with their known structures in the published literature \citep{Bhat2018,Xue2017}. Also notable is PSR J1900$-$2600, which is known to exhibit a dramatic evolution in the shape of its pulse profile and a complex, multi-component structure at higher frequencies \citep{Mitra2008}. The AAVS2 and EDA2 detections of this pulsar show good agreement with the profile published by \cite{Xue2017} from MWA observations at \SI{185}{\MHz}.

A closer examination of the European Pulsar Network (EPN) Database\footnote{\url{http://www.epta.eu.org/epndb/}} and studies of the pulsar population at low frequencies \citep[e.g.][]{Bon2020,Xue2017,Stovall2015} reveals that only a small fraction of pulsars have published detections below \SI{400}{\MHz}, and therefore detection of even a modest sample can make a useful contribution to the growing database. In fact, six detections in this paper were made at the lowest frequencies published to date: PSRs J0835$-$4510\footnote{The continuum flux density measurements from \citet{Murphy2017} and \citet{Bell2016} are excluded as they are not beamformed detections.} (\SI{159.4}{\MHz}), J1116$-$4122 (\SI{150}{\MHz}), J1453$-$6413 (\SI{101.6}{\MHz}), J1456$-$6843 (\SI{101.6}{\MHz}), J1731$-$4744 (\SI{132.8}{\MHz}), and J1751$-$4657 (\SI{150}{\MHz}). Each of these pulsars were also detected by \citet{Xue2017} at \SI{185}{\MHz} using incoherently-summed MWA data, however the SKA-Low station detections were made down to even lower frequencies, in addition to having coherently summed station beams. Moreover, the detections of the Vela pulsar (PSR J0835$-$4510) are marginally lower in frequency than the MWA detections published in \cite{Kirsten2019} at \SI{164.5}{\MHz}.

\subsection{Mean flux densities}
\label{subsec:flux}
\begin{figure}[t]
    \centering
    \includegraphics[width=\linewidth]{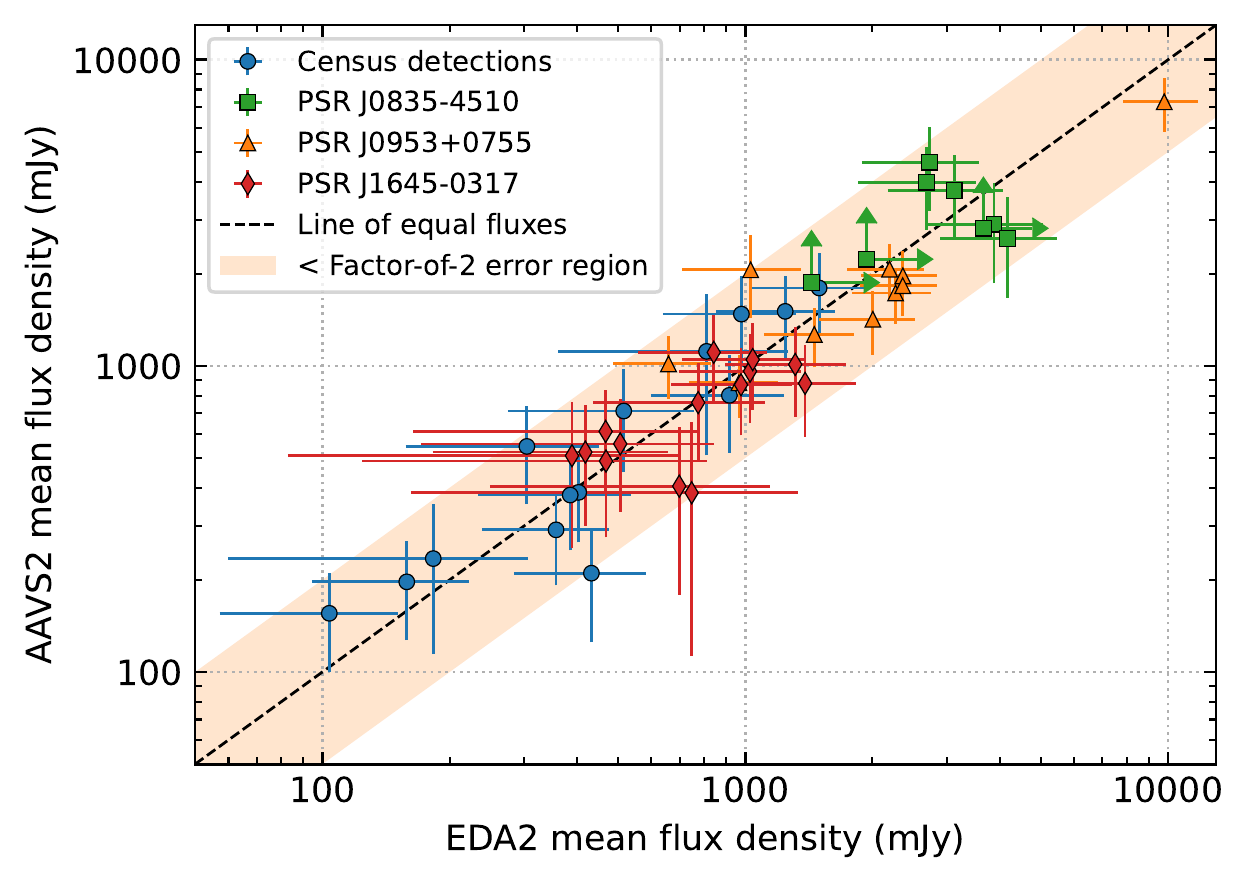}
    \caption{Comparison of the mean flux density measurements obtained from observations performed with the EDA2 and AAVS2 stations. The data points represent measurements from simultaneous detections with the two stations. Blue filled circles: detections of 19 pulsars detected at single frequencies in the preliminary shallow all-sky census. Green squares: detections of PSR J0835$-$4510 made between 148--\SI{352}{\MHz} (lower limits are indicated with green arrows). Orange triangles: detections of PSR J0953+0755 made between 86--\SI{328}{\MHz}. Red diamonds: detections of PSR J1645$-$0317 made between 86--\SI{352}{\MHz}. Black dashed line: the line of equal fluxes. Orange shaded envelope: the region between 50\% and 200\% of the equal flux value.}
    \label{fig:fluxcompare}
\end{figure}
Mean flux densities were measured for 22 pulsars using the methods described in \S\ref{subsec:calcflux} and \S\ref{subsec:cal}. The results are given in \tab\ref{tab:detections}. For eight pulsars, detections were made at multiple frequencies and mean flux densities were measured for each detection. Multiple simultaneous detections were made of PSRs J0835$-$4510, J0953+0755 and J1645$-$0317. Additionally, most of the pulsar detections from the initial census were made with both stations. There is therefore a substantial number of simultaneous observations which can be used to independently verify the accuracy of the mean flux density measurements. \fig\ref{fig:fluxcompare} shows a comparison of the mean flux densities at frequencies for which observations were performed with both stations simultaneously. Nearly all measurements agree within a factor of 2, despite the stations using different antenna types and relying on beam models with different accuracies. The plot also indicates the sensitivities of the stations, with detections being made down to $\sim100\mJy$ (for a $\sim10\sigma$ detection of the pulsar in 30$\,$min, at \SI{150}{\MHz}). The agreement between the stations gives us confidence that the mean flux density measurements are reasonably accurate and that their calculated uncertainties are not underestimated.

\begin{table*}[ht]
    \begin{threeparttable}[c]
        \caption{Measured flux densities of the detected pulsars and parameters of the observations performed with the EDA2 and AAVS2 stations.}
        \label{tab:detections}
        \begin{tabular}[c]{lccccccccc}
        	\toprule
        	Pulsar name & Period\tnote{a} & DM\tnote{b}  & $\nu$\tnote{c} & T\textsubscript{AAVS2}\tnote{d} & T\textsubscript{EDA2}\tnote{d} & S/N\textsubscript{AAVS2}\tnote{e} & S/N\textsubscript{EDA2}\tnote{e} & $S_{\nu,\text{AAVS2}}$\tnote{f} & $S_{\nu,\text{EDA2}}$\tnote{f} \\
        	 & (ms) & (\SI{}{\centi\metre\tothe{-3}} pc) & (MHz) & (min) & (min) & & & (Jy) & (Jy) \\
            \midrule
            J0034$-$0721 & 942.95 & 10.92 & 79.7 & 30 & -- & 12 & -- & 0.7 $\pm$ 0.2 & --\\
            J0437$-$4715\tnote{$\star$} & 5.76 & 2.64 & 150.0 & 30 & 30 & 32 & 21 & 0.4 $\pm$ 0.1 & 0.4 $\pm$ 0.1\\
            J0452$-$1759 & 548.94 & 39.9 & 150.0 & 30 & 30 & 9 & 8 & 0.16 $\pm$ 0.06 & 0.10 $\pm$ 0.05\\
            J0630$-$2834 & 1244.42 & 34.42 & 79.7 & 30 & 30 & 102 & 36 & 1.8 $\pm$ 0.5 & 1.5 $\pm$ 0.5\\
            J0820$-$1350 & 1238.13 & 40.94 & 79.7 & 30 & 30 & 11 & 8 & 0.4 $\pm$ 0.1 & 0.4 $\pm$ 0.2\\
            J0826+2637 & 530.66 & 19.48 & 100.0 & 30 & 30 & 9 & 11 & 0.7 $\pm$ 0.3 & 0.5 $\pm$ 0.2\\
            J0835$-$4510 & 89.33 & 67.77 & 320.3 & 30 & 30 & 195 & 21 & 4 $\pm$ 1 & 2.7 $\pm$ 0.8\\
            J0837+0610 & 1273.77 & 12.86 & 79.7 & 30 & 30 & 80 & 19 & 1.5 $\pm$ 0.5 & 1.2 $\pm$ 0.4\\
            J0953+0755 & 253.07 & 2.97 & 150.0 & 15 & 20 & 50 & 30 & 2.1 $\pm$ 0.6 & 1.0 $\pm$ 0.3\\
            J1116$-$4122 & 943.16 & 40.53 & 150.0 & 30 & 30 & 16 & 11 & 0.20 $\pm$ 0.07 & 0.16 $\pm$ 0.06\\
            J1453$-$6413 & 179.49 & 71.25 & 200.0 & 30 & 30 & 16 & 18 & 0.8 $\pm$ 0.3 & 0.9 $\pm$ 0.3\\
            J1456$-$6843 & 263.38 & 8.61 & 150.0 & 30 & 30 & 19 & 13 & 1.5 $\pm$ 0.5 & 1.0 $\pm$ 0.3\\
            J1645$-$0317 & 387.69 & 35.76 & 200.0 & 30 & 30 & 23 & 53 & 0.9 $\pm$ 0.3 & 1.0 $\pm$ 0.3\\
            J1709$-$1640 & 653.05 & 24.89 & 79.7 & 30 & -- & 7 & -- & 0.6 $\pm$ 0.3 & --\\
            J1731$-$4744 & 829.83 & 123.06 & 150.0 & 30 & 30 & 16 & 8 & 0.5 $\pm$ 0.2 & 0.3 $\pm$ 0.1\\
            J1751$-$4657 & 742.35 & 20.4 & 150.0 & 30 & -- & 8 & -- & 0.3 $\pm$ 0.1 & --\\
            J1820$-$0427 & 598.08 & 84.44 & 200.0 & -- & 30 & -- & 16 & -- & 0.4 $\pm$ 0.2\\
            J1900$-$2600 & 612.21 & 37.99 & 200.0 & 30 & 15 & 9 & 8 & 0.21 $\pm$ 0.09 & 0.4 $\pm$ 0.1\\
            J1913$-$0440 & 825.94 & 89.38 & 150.0 & 30 & 30 & 5 & 14 & 0.2 $\pm$ 0.1 & 0.2 $\pm$ 0.1\\
            J1932+1059 & 226.52 & 3.18 & 150.0 & -- & 30 & -- & 4 & -- & 0.3 $\pm$ 0.2\\
            J2018+2839 & 557.95 & 14.2 & 150.0 & 30 & 30 & 8 & 4 & 1.1 $\pm$ 0.6 & 0.8 $\pm$ 0.4\\
            J2048$-$1616 & 1961.57 & 11.46 & 150.0 & 30 & 30 & 26 & 12 & 0.3 $\pm$ 0.1 & 0.4 $\pm$ 0.1\\
        	\bottomrule
        \end{tabular}
        \begin{tablenotes}
            \item [a] Pulsar period listed in the ATNF pulsar catalogue.
            \item [b] DM listed in the ATNF pulsar catalogue.
        	\item [c] Centre frequency of the observation.
        	\item [d] Observation duration.
        	\item [e] S/N of the average pulse profile obtained from EDA2 and AAVS2 data.
        	\item [f] Mean flux density calculated from the average pulse profile obtained from EDA2 and AAVS2 data.
        	\item [$\star$] Millisecond pulsar.
        \end{tablenotes}
    \end{threeparttable}
\end{table*}

\begin{table*}[hp]
    \begin{threeparttable}
        \caption{Flux density measurements at three example frequencies and best-fitting model parameters for the pulsars observed at multiple frequencies using the SKA-Low stations.}
        \label{tab:fit}
        \begin{tabular}{lcccccccccc}
        	\toprule
            Pulsar name & $p_\mathrm{best}$\tnote{a} & $S_{102}$\tnote{b} & $S_{180}$\tnote{b} & $S_{289}$\tnote{b} & $a_0$\tnote{c} & $a_1$\tnote{d} & $a_2$\tnote{e} & Type\tnote{f} & Fit Range\tnote{g} & Notes \\
            & & (Jy) & (Jy) & (Jy) & & & & & (MHz) & \\
            \midrule
            J0820$-$1350 & 0.99 & -- & 0.3 $\pm$ 0.1 & -- & 440 $\pm$ 60 & $-$0.83 $\pm$ 0.08 & $-$2.2 $\pm$ 0.2 & BPL & 59--4820 & \\
            J0835$-$4510 & 1.00 & -- & 3.3 $\pm$ 0.7 & 2.6 $\pm$ 0.8 & 970 $\pm$ 50 & $-$0.50 $\pm$ 0.03 & $-$2.9 $\pm$ 0.2 & BPL & 76--343500 &  L\\
            J0837+0610 & 0.81 & 0.4 $\pm$ 0.1 & 0.3 $\pm$ 0.1 & 0.3 $\pm$ 0.2 & 65 $\pm$ 6 & $-$1.34 $\pm$ 0.09 & $-$1.55 $\pm$ 0.05 & LPS & 20--3100 &  U,E\\
            J0953+0755 & 1.00 & 2.0 $\pm$ 0.3 & 2.1 $\pm$ 0.3 & 4.1 $\pm$ 0.6 & 104 $\pm$ 7 & $-$2.6 $\pm$ 0.2 & 0.53 $\pm$ 0.09 & PL-T & 20--22700 & \\
            J1453$-$6413 & 0.68 & 0.9 $\pm$ 0.4 & 0.8 $\pm$ 0.2 & 0.3 $\pm$ 0.2 & 140 $\pm$ 40 & $-$2.5 $\pm$ 0.3 & 1.4 $\pm$ 0.8 & PL-T & 102--8400 &  U,E\\
            J1456$-$6843 & 0.42 & 1.0 $\pm$ 0.4 & 1.4 $\pm$ 0.5 & -- & 7 $\pm$ 19 & $-$0.3 $\pm$ 0.2 & $-$1.4 $\pm$ 0.1 & LPS & 102--8400 &  N\\
            J1645$-$0317 & 0.98 & 0.5 $\pm$ 0.2 & 1.1 $\pm$ 0.4 & 0.8 $\pm$ 0.2 & 276 $\pm$ 2 & 0.22 $\pm$ 0.01 & $-$2.15 $\pm$ 0.01 & BPL & 50--22700 &  N\\
            J1731$-$4744 & 0.60 & -- & 0.5 $\pm$ 0.2 & -- & $-$1.38 $\pm$ 0.05 & -- & -- & PL & 133--3100 &  U\\
            \bottomrule
        \end{tabular}
        \begin{tablenotes}
            \item [a] Probability that the best-fitting model is the true best-fitting model.
            \item [b] Flux density calculated from EDA2 and AAVS2 detections. For frequencies where detections were made with both stations, a weighted average is given.
            \item [c] The spectral index (PL), the break frequency in MHz (BPL), the turn-over frequency in MHz (PL-T), the peak frequency in MHz (LPS), or the cut-off frequency in MHz (PL-C).
            \item [d] The spectral index before the break (BPL), the spectral index (PL-T and PL-C), or the curvature parameter (LPS).
            \item [e] The spectral index after the break (BPL), the smoothness parameter (PL-T), or the spectral index for the case of zero curvature (LPS).
            \item [f] The spectral classification. PL: simple power-law. BPL: broken power-law. LPS: log-parabolic spectrum. PL-T: power-law with low-frequency turn-over. PL-C: power-law with high-frequency cut-off.
            \item [g] The frequency range of the fitted data set.
            \item [L] Due to the significant scattering tail of PSR J0835-4510, the flux densities below \SI{200}{\MHz} were calibrated using upper noise limits, and may be underestimates (see text for details).
            \item [E] Pulsars whose flux densities were calibrated against full electromagnetic simulations.
            \item[U] Best-fitting model updated from \citetalias{Jankowski2018}.
            \item[N] Not modelled by \citetalias{Jankowski2018}.
        \end{tablenotes}
    \end{threeparttable}
\end{table*}

\begin{table*}[hp]
    \begin{threeparttable}
        \caption{Flux density measurements and best-fitting model parameters for the pulsars observed at single frequencies using the SKA-Low stations.}
        \label{tab:fit_spot}
        \begin{tabular}{lccccccccc}
        	\toprule
            Pulsar name & $p_\mathrm{{best}}$\tnote{a} & $\nu$\tnote{b} & $S_\nu$\tnote{c} & $a_0$\tnote{a} & $a_1$\tnote{a} & $a_2$\tnote{a} & Type\tnote{a} & Fit Range\tnote{a} & Notes \\
            & & (MHz) & (Jy) & & & & & (MHz) & \\
            \midrule
            J0034$-$0721 & 0.63 & 79.7 & 0.7 $\pm$ 0.2 & 55 $\pm$ 5 & $-$3 $\pm$ 1 & 0.6 $\pm$ 0.2 & PL-T & 20--2388 &  N\\
            J0437$-$4715 & 1.00 & 150.0 & 0.40 $\pm$ 0.09 & 1900 $\pm$ 100 & $-$0.830 $\pm$ 0.009 & $-$2.4 $\pm$ 0.2 & BPL & 76--17000 & \\
            J0452$-$1759 & 1.00 & 150.0 & 0.13 $\pm$ 0.04 & 1490 $\pm$ 20 & $-$0.1 $\pm$ 0.1 & -- & PL-C & 102--1408 &  N\\
            J0630$-$2834 & 1.00 & 79.7 & 1.7 $\pm$ 0.4 & 98 $\pm$ 4 & $-$2.5 $\pm$ 0.2 & 2.1 $\pm$ 0.1 & PL-T & 35--8400 & \\
            J0826+2637 & 0.99 & 100.0 & 0.6 $\pm$ 0.2 & 160 $\pm$ 10 & 0.3 $\pm$ 0.1 & $-$1.62 $\pm$ 0.03 & BPL & 20--22700 &  N\\
            J1116$-$4122 & 0.98 & 150.0 & 0.18 $\pm$ 0.05 & 3400 $\pm$ 200 & $-$1.36 $\pm$ 0.08 & -- & PL-C & 150--3100 &  U\\
            J1709$-$1640 & 0.86 & 79.7 & 0.6 $\pm$ 0.3 & 70 $\pm$ 10 & $-$1.7 $\pm$ 0.3 & 2 $\pm$ 1 & PL-T & 50--22700 &  N\\
            J1751$-$4657 & 0.55 & 150.0 & 0.3 $\pm$ 0.1 & 48 $\pm$ 54 & $-$1.0 $\pm$ 0.5 & $-$2.3 $\pm$ 0.2 & LPS & 150--3100 &  U\\
            J1820$-$0427 & 0.81 & 200.0 & 0.4 $\pm$ 0.2 & 220 $\pm$ 2 & 1.73 $\pm$ 0.03 & $-$2.21 $\pm$ 0.02 & BPL & 102--4920 &  N\\
            J1900$-$2600 & 0.52 & 200.0 & 0.4 $\pm$ 0.1 & 15 $\pm$ 12 & $-$0.5 $\pm$ 0.1 & $-$1.86 $\pm$ 0.07 & LPS & 154--4820 &  U\\
            J1913$-$0440 & 0.86 & 150.0 & 0.21 $\pm$ 0.09 & 150 $\pm$ 28 & $-$1.7 $\pm$ 0.3 & $-$1.9 $\pm$ 0.1 & LPS & 102--3100 &  U\\
            J1932+1059 & 1.00 & 150.0 & 0.3 $\pm$ 0.2 & 90 $\pm$ 10 & $-$1.5 $\pm$ 0.2 & 0.9 $\pm$ 0.2 & PL-T & 20--22700 &  N\\
            J2018+2839 & 0.96 & 150.0 & 1.0 $\pm$ 0.4 & 240 $\pm$ 20 & 1.0 $\pm$ 0.2 & $-$2.1 $\pm$ 0.1 & BPL & 25--22700 &  N\\
            J2048$-$1616 & 0.73 & 150.0 & 0.33 $\pm$ 0.08 & 180 $\pm$ 40 & $-$1.1 $\pm$ 0.1 & 2 $\pm$ 2 & PL-T & 102--22700 &  U\\
            \bottomrule                               
        \end{tabular}
        \begin{tablenotes}
            \item [a] See notes of Table \ref{tab:fit} for details.
            \item [b] Centre frequency of the observation. These frequencies were chosen based on where the S/N was expected to be highest.
            \item [c] Flux density calculated from EDA2 and AAVS2 detections at centre frequency $\nu$. For frequencies where detections were made with both stations, a weighted average is given.
            \item[U] Best-fitting model updated from \citetalias{Jankowski2018}.
            \item[N] Not modelled by \citetalias{Jankowski2018}.
        \end{tablenotes}
    \end{threeparttable}
\end{table*}

\subsection{Flux density spectra}
\label{subsec:specmod}

\begin{figure*}[!ht]
    \centering
    \includegraphics[width=\linewidth]{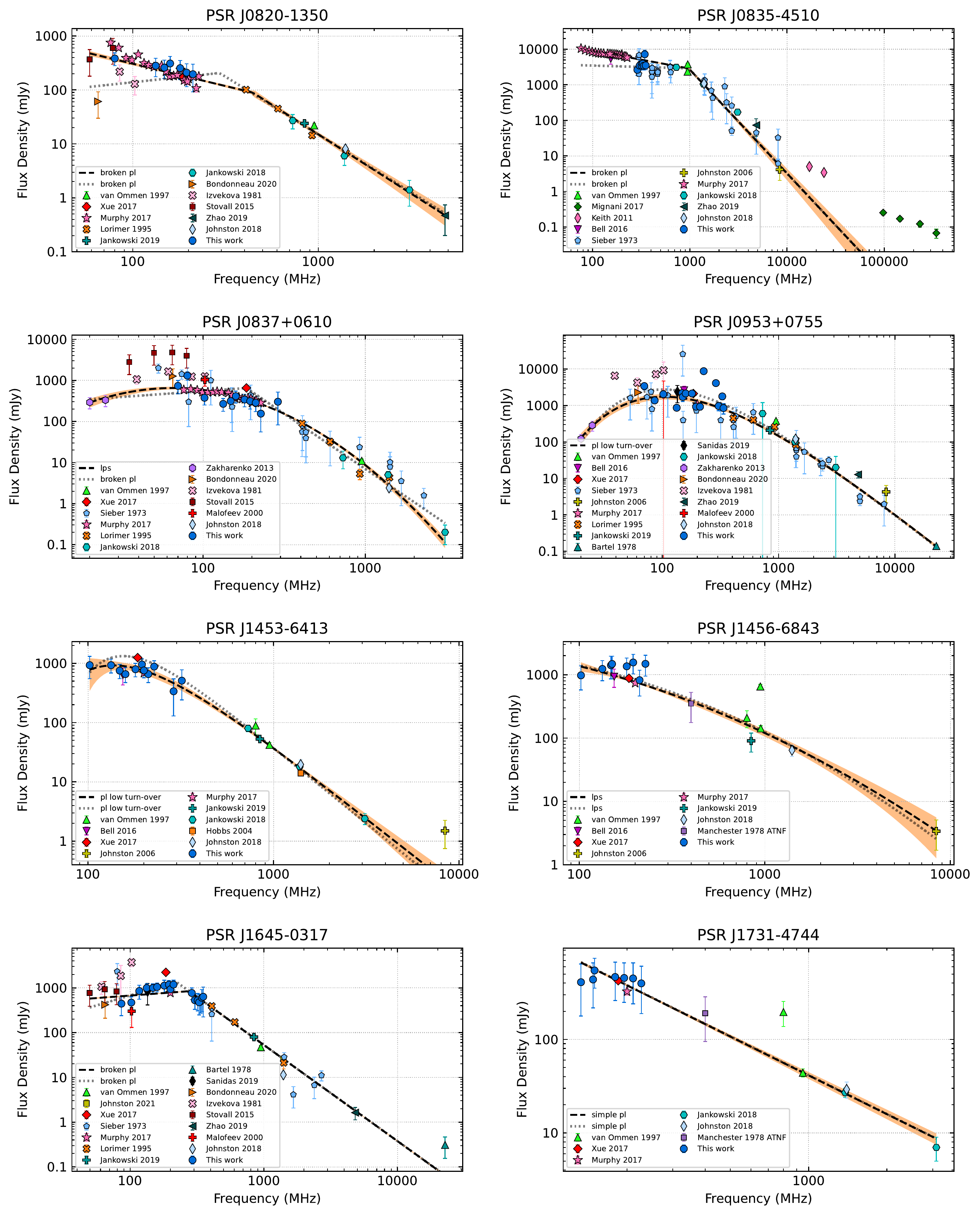}
    \caption{Flux density spectra for the 8 pulsars whose mean flux densities were measured at multiple frequencies using the EDA2 and AAVS2 stations. Black dashed line: the best-fitting model to the data. Orange shaded envelope: the 1$\sigma$ uncertainty of the best-fitting model. Grey dotted line: the best-fitting model to the data when continuum flux density measurements are excluded from the fit.}
    \label{fig:spec1}
\end{figure*}

The spectra of all 22 pulsars which were detected with the SKA-Low stations were modelled using the method described in \S\ref{subsec:mod}. The data sets for each spectrum had a sufficient number of points to fit all model types and spanned a sufficiently wide frequency range to make a meaningful fit. The results for the pulsars which were observed at multiple frequencies are presented in \tab\ref{tab:fit} and \fig\ref{fig:spec1}, and each of these fits are analysed in detail in \S\ref{sec:pmf}. For the pulsars which were only observed at single frequencies, the results are given in \tab\ref{tab:fit_spot} and \fig\ref{fig:spec2}.

The most common spectral types were power-law with low-frequency turn-over and broken power-law, with seven pulsars being best fit by each of these models, followed by five pulsars exhibiting a log-parabolic spectrum. The power-law with high-frequency hard cut-off was seen to be exhibited by two pulsars, and the simple power-law model by one pulsar. Of the 22 analysed pulsars, nine were not previously classified in \citetalias{Jankowski2018}, and for eight we have updated the best-fitting models. These pulsars are indicated in Tables \ref{tab:fit} and \ref{tab:fit_spot}. As a result of the current sensitivity limitation of SKA-Low stations, our detections are primarily for brighter pulsars, most of which have some low-frequency measurements available. Furthermore, all but one pulsar (J1731$-$4744) show deviations from a simple power-law model. This suggests that pulsars with well-determined spectra are more likely to show spectral features, which is in line with the findings from \citetalias{Jankowski2018}.

\begin{figure*}[!ht]
    \centering
    \includegraphics[width=\linewidth]{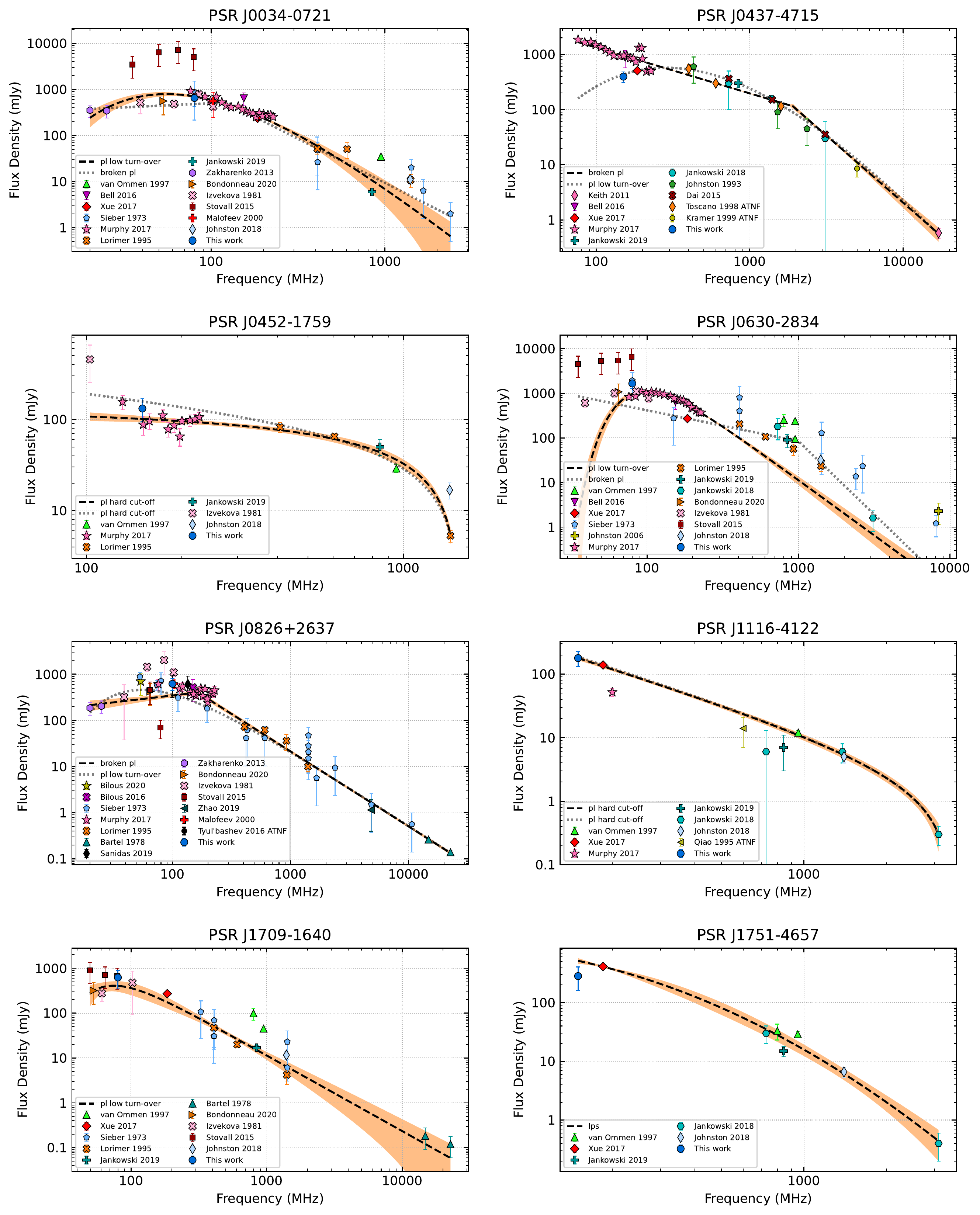}
    \caption{Flux density spectra for the 14 pulsars whose mean flux densities were measured at single frequencies using the EDA2 and AAVS2 stations. Black dashed line: the best-fitting model to the data. Orange shaded envelope: the 1$\sigma$ uncertainty of the best-fitting model. Grey dotted line: the best-fitting model to the data when continuum flux density measurements are excluded from the fit.}
    \label{fig:spec2}
\end{figure*}
\begin{figure*}[!ht]
    \continuedfloat
    \centering
    \includegraphics[width=\linewidth]{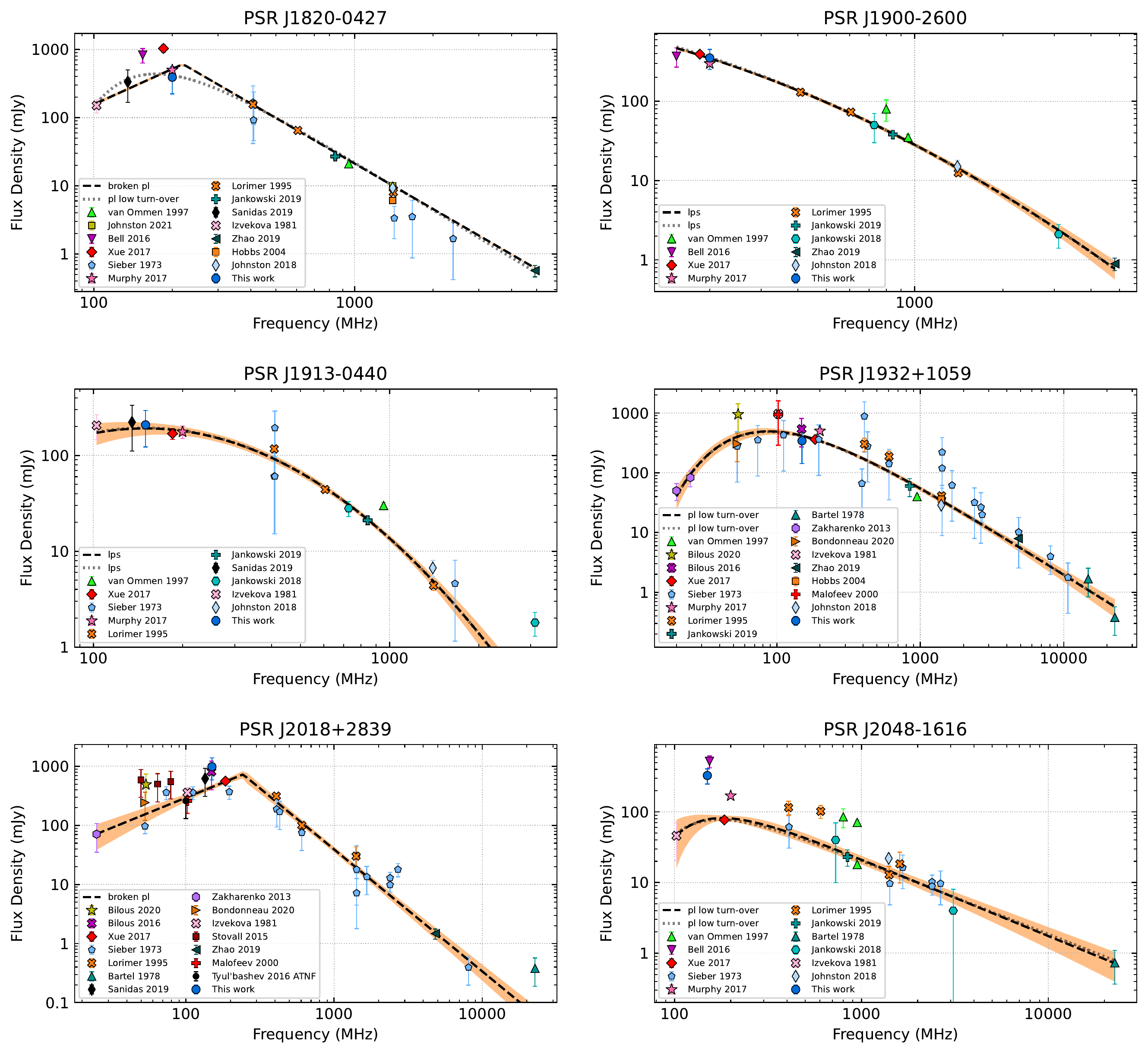}
    \caption{continued.}
\end{figure*}

\subsubsection{Modelling without continuum flux densities}
\label{subsubsec:imaging}
Measurements of flux density from continuum images can have much lower measurement uncertainties than traditional flux density measurements from beamformed detections if the field-of-view is large enough to enable robust in-field calibration to be performed. However, continuum measurements can be vulnerable to systematic errors caused by blending with emission from other radio sources within the beam. It can therefore be difficult to determine the intrinsic flux of the pulsar \citep[e.g.][]{Bell2016}. Observations of targets near the Galactic plane ($|b|\lesssim5\degree$) are more likely to be affected by source confusion, and the problem may be amplified for larger beam sizes, such as the $\sim3\,$arcmin MWA beam used for the GLEAM survey \citep{Wayth2015}.

Only five of the detected pulsars lie on the Galactic plane: J0835$-$4510, J1453$-$6413, J1820$-$0427, J1932+1059, J2018+2839. Of these pulsars, none show notable discrepancies between continuum measurements and our measurements. Nevertheless, for the spectral modelling presented in Figures \ref{fig:spec1} and \ref{fig:spec2}, we have considered flux densities measured from pulsed emission and continuum images as two separate classes of measurements. In addition to the models fitted to the full data sets, secondary models were fitted without the contributions from continuum flux density measurements, which includes all measurements from \cite{Murphy2017} and \cite{Bell2016} (i.e. the secondary fits excluded data from these two publications). For the most part, the two data sets suggested similar models, with a few exceptions which are discussed in \S\ref{sec:pmf}.

\subsubsection{Analysis of pulsar spectra}\label{sec:pmf}

\paragraph{PSR J0437$-$4715:} Being the only MSP in the sample, the spectral behaviour of this pulsar is of particular interest. Our modelling shows that the best-fitting model is a broken power-law with a spectral break at \measu{1900}{100}{MHz} and a steep spectral index of \meas{$-$2.4}{0.2} above the break frequency. This is in agreement with \citetalias{Jankowski2018}, and lowers the uncertainties of these two parameters significantly. Interestingly, our analysis suggests that the spectrum of this pulsar shows a low-frequency turn-over at \measu{285}{5}{MHz} when continuum flux densities are excluded from the fit. It is important to note that MSPs are generally not seen to turn over, and those that hint at a turn-over are expected to peak well below \SI{100}{\MHz}. Furthermore, this pulsar shows high variability at low frequencies due to interstellar scintillation. \cite{Bhat2018} measured a characteristic scintillation bandwidth of \measu{1.4}{0.4}{MHz}, which makes the SKA-Low stations susceptible to intensity variations between observations. This may explain the discrepancy between the EDA2 measurements and the \citet{Murphy2017} measurements (around a factor of 2).

\paragraph{PSR J0630$-$2834:} This pulsar shows perhaps the most striking example of continuum measurements dominating the spectral fit. The primary model fit indicates that the pulsar turns over at \measu{98}{4}{MHz}, however the model departs significantly from the majority of data points, with the exception of the \cite{Murphy2017} data. Excluding continuum measurements, we find that the best-fitting model is a broken power-law with a break frequency of \measu{1000}{200}{MHz}, which tracks more closely with the majority of the available data. Evidently, this pulsar would benefit from more low-frequency measurements to clarify these inconsistencies.

\paragraph{PSR J0820$-$1350:} This pulsar has a well-defined spectrum, showing a power-law with a break frequency of \measu{440}{60}{MHz}, which is in agreement with \citetalias{Jankowski2018} and reduces the uncertainty by 40\%. Our modelling also shows shallower power-laws on either side of the break frequency than those found in \citetalias{Jankowski2018}, although the values agree within their respective uncertainties. Interestingly, the exclusion of imaging measurements leaves the spectrum poorly constrained below \SI{400}{\MHz}, leading to the broken power-law peaking at \measu{300}{40}{MHz}, with a positive spectral index below the break.

\paragraph{PSR J0835$-$4510:} Although the Vela pulsar is one of the brightest and most well-studied pulsars, if we limit ourselves to traditional (beamformed) pulsar observations, its radio spectrum remains relatively sparsely populated and poorly constrained at low frequencies. Low-frequency measurements are limited by interstellar scattering, which leads to substantial temporal broadening at frequencies below \SI{300}{\MHz} and renders pulses undetectable below $\sim\SI{150}{\MHz}$. At these frequencies, the only measurements available are from continuum imaging. At higher frequencies (between \SI{600}{\MHz} and \SI{10}{\GHz}), the only strong constraints are provided by \citetalias{Jankowski2018}, \citet{Johnston2018}, \citet{Zhao2019}, and \citet{vanOmmen1997}. Above \SI{10}{\GHz}, the model departs significantly from the data. Notably, this fit includes measurements from \citet{Mignani2017} at millimetre and even sub-millimetre wavelengths made using the Atacama Large Millimeter Array (ALMA). Very few pulsars have been observed at these extremely high frequencies, and it is likely that the models considered in this work do not adequately describe the behaviour of the full spectrum. A more complex model such as a double broken power-law \cite[see, for example,][]{Bilous2016} may be more suitable. Notwithstanding, our modelling shows a broken power-law with a spectral break at \measu{970}{50}{MHz} and a spectral index below the break of \meas{$-$0.50}{0.03}, which are in agreement with \citetalias{Jankowski2018}. Above the break, our modelling suggests a steeper spectrum than reported in \citetalias{Jankowski2018}, with a spectral index of \meas{$-$2.9}{0.2}.

\paragraph{PSR J0837+0610:} This pulsar has a substantial amount of data available over a moderately wide frequency range. Our modelling shows that the best-fitting model is a log-parabolic spectrum that peaks at \measu{65}{6}{MHz}. It is clear that the measurements from \cite{Murphy2017} dominate the spectral fit, and that the reliability of the model fit is heavily dependent upon these continuum measurements. Our flux density measurements fall in line with these measurements, barring some variability resulting from interstellar scintillation, an effect typically seen in low-DM pulsars such as this ($\text{DM}\approx 12.86\dmu$). Nevertheless, excluding continuum measurements, we find that the spectrum exhibits a broken power-law with a break at \measu{185.00}{0.04}{MHz}, constrained strongly by a low-uncertainty measurement from \citet{Xue2017}.

\paragraph{PSR J0953+0755:} Being one of the nearest and brightest pulsars, J0953+0755 has one of the best-determined radio spectra, with data available from 17 publications in our database and spanning three orders of magnitude in frequency. Our modelling shows that this pulsar exhibits a low-frequency turn-over at \measu{104}{7}{MHz}, which is marginally higher than that which was found in \citetalias{Jankowski2018}. This pulsar is also known to exhibit significant variability \citep[see][]{Bell2016}, due to which measurements based on single observations may ideally need to account for additional sources of errors (typically a factor of $\sim2$--3 at these frequencies). It is likely that scintillation is also the cause of the variability seen in measurements from \cite{Sieber1973} and \cite{Izvekova1981} at frequencies below $\sim\SI{1}{\GHz}$. For this spectrum, the contribution from continuum measurements is minimal.

\paragraph{PSR J1453$-$6413:} The best-fitting model for this pulsar is a steep power-law with a spectral index of \meas{$-$2.5}{0.3} and a low-frequency turn-over at \measu{140}{40}{MHz}. For this pulsar, our measurements are the lowest frequency available, and thus important in constraining the turn-over. This pulsar was previously found to have a broken power-law with a break at \measu{320}{30}{MHz} \citepalias{Jankowski2018}, however our data suggest that there may be significant curvature below $\SI{300}{\MHz}$.

\paragraph{PSR J1456$-$6843:} For this pulsar, we found that the best-fitting model is a log-parabolic spectrum with a peak at $\lesssim\SI{30}{\MHz}$, however it is unlikely that the spectrum truly peaks at such a low frequency. The far southern declination of the pulsar is perhaps the most likely reason for the relatively small number of measurements. More data would be beneficial to better constrain the spectrum.

\paragraph{PSR J1645$-$0317:} Our modelling shows that this pulsar exhibits a broken power-law with a peak at \measu{276}{2}{MHz} and a shallow positive spectral index of \meas{0.22}{0.01} below the break. Above the break, the power-law has a spectral index of \meas{$-$2.15}{0.01}, which is shallower than the spectral index of \meas{$-$2.6}{0.2} reported in \citet{Johnston2021}. Excluding the continuum measurement from \citet{Murphy2017}, the peak of the broken power-law shifts to \measu{220}{2}{MHz}, with a steeper spectral index of \meas{0.89}{0.01} below the break. Our data fills a large gap in the spectrum of this pulsar and plays an important role in constraining its spectrum.

\paragraph{PSR J1731$-$4744:} Our observations of this pulsar have yielded the lowest-frequency measurements in the spectrum, and our analysis hints at flattening or turn-over at $\lesssim\SI{200}{\MHz}$. However, the large uncertainties of the measurements mean limited constraints on the model, which we find to be a simple power-law with a relatively shallow spectral index of \meas{$-$1.38}{0.05}.

\begin{table}[t]
    \begin{threeparttable}
        \caption{Estimated magnetic field strengths and emission heights based on the power-law with high-frequency cut-off model.}
        \label{tab:emission_heights}
        \begin{tabular}{lcc}
        	\toprule
            Pulsar name & J0452$-$1759 & J1116$-$4122 \\
            \midrule
            $B_\mathrm{surf}$ ($10^{12}\Gauss$)\tnote{a} & 1.80 & 2.77 \\
            $B_\mathrm{LC}$ (G)\tnote{a} & 101.9 & 31.0 \\
            $B_\mathrm{pc}$ ($10^{11}\Gauss$) & \meas{0.221}{0.006} & \meas{2.0}{0.2} \\
            $z_e$ (km) & \meas{52}{9} & \meas{29}{5} \\
            $z_e/R_\text{LC}$ (\%) & \meas{0.20}{0.03} & \meas{0.06}{0.01} \\
            \bottomrule                                     
        \end{tabular}
        \begin{tablenotes}
            \item [a] Values listed in the ATNF pulsar catalogue.
        \end{tablenotes}
    \end{threeparttable}
\end{table}

\paragraph{PSRs J0452$-$1759 and J1116-4122:} These are the only pulsars in our sample to show a high-frequency hard cut-off, a model proposed by \citet{Kontorovich2013} to describe the coherent emission of electrons when accelerated by the pulsar's electric field. Following \citetalias{Jankowski2018}, we calculate the magnetic field strength in the centre of the polar cap as
\begin{equation}
    B_\text{pc} = \frac{m_e c}{\pi e}P\nu_c^2,
\end{equation}
where $m_e$ and $e$ are the mass and charge of the electron, $c$ is the speed of light, $P$ is the pulsar period, and $\nu_c$ is the cut-off frequency obtained from the spectral fit. If we assume a dipolar magnetic field, the field strength drops off as $z^{-3}$, where $z$ is the distance from the centre of the neutron star. We can thus derive the emission height (i.e. the height of the centre of the polar cap) $z_e$ as
\begin{equation}
    z_e = z_\text{surf}\qty(\frac{B_\text{pc}}{B_\text{surf}})^{-1/3},
\end{equation}
where $z_\text{surf}$ is the radius of a canonical 1.4$\,\text{M}_\odot$ neutron star \citep[\measu{12}{2}{km};][]{Steiner2018} and $B_\text{surf}$ is the magnetic field strength at the surface of the star. Using the cut-off frequencies given in \tab\ref{tab:fit_spot}, we calculated $B_\text{pc}$ and $z_e$ for both pulsars. The results are presented in \tab\ref{tab:emission_heights}, along with the magnetic field strength at the light-cylinder radius $B_\mathrm{LC}$ and the emission height normalised by the light-cylinder radius $z_e/R_\mathrm{LC}$. Our fits predict $z_e\sim30$--50$\,$km, which is in line with \citetalias{Jankowski2018}, who noted that the estimated emission heights are unreasonably low and that further constraints will be required at high frequencies to test the validity of the model and its underlying assumptions. Furthermore, for PSR J0452$-$1759, our fitted spectral index of \meas{$-$0.1}{0.1} departs significantly from the assumed $\alpha=-2$ in \citet{Kontorovich2013}. More data would be beneficial to better constrain this spectrum and clarify this inconsistency.

\section{SUMMARY AND CONCLUSIONS}
\label{sec:con}
We have presented the first pulsar detections with the second-generation SKA-Low precursor stations, the AAVS2 and the EDA2. By performing a shallow all-sky census of the southern sky, we have detected 22 bright known pulsars at low frequencies. With the current modest station sensitivities ($\sim\SI{1}{\MHz}$ bandwidth and 30$\,$min integration time), the stations are sensitive down to $\sim100\mJy$ for pulsar detections. Furthermore, follow-up observations were made across a wide frequency range ($\sim70$--\SI{350}{MHz}), yielding detections at multiple frequencies for eight pulsars. Six pulsars observed in this work were detected at the lowest frequencies ever published (J0835$-$4510, J1116$-$4122, J1453$-$6413, J1456$-$6843, J1731$-$4744, J1751$-$4657). In general, the AAVS2 produced higher-S/N pulsar detections than the EDA2, which is likely a result of the differences in antenna types between the stations. We expect that the upcoming system bandwidth upgrade ($\gtrsim\SI{25}{\MHz}$) will enable a five-fold increase in sensitivity, and yield high-quality pulsar detections with only a fraction of the processing time of the MWA, making the stations strategically useful tools for regular pulsar monitoring.

Flux densities were measured for all 22 detected pulsars, which show close agreement between the two stations and with other measurements in literature. This gives us confidence that the sensitivity simulation code used to calibrate the observations produces consistently reliable estimates. Comparisons with full-electromagnetic simulations confirm this, with a typical discrepancy of no more than 30\%.

We have revisited the spectral modelling of 13 pulsars previously analysed in \citetalias{Jankowski2018}, and further analysed an additional 9 by employing the robust modelling and classification method described in the original publication. By augmenting our results with a compilation of flux density measurements from the published literature, we have classified the analysed spectra into 5 morphological classes. For 17 pulsars, we have presented new or updated best-fitting models. In our sample of bright pulsars, all but one spectrum showed deviations from simple power-law behaviour, which suggests that populating the low-frequency spectra of pulsars will often reveal hidden spectral features. Furthermore, we have found that even very well-studied pulsars such as the Vela pulsar (PSR J0835$-$4510) often have poorly constrained spectra and could benefit from further measurements to fill gaps in the data and clarify inconsistencies between published values. Thus, it is important for more pulsar flux density measurements to be made at low frequencies, ideally with instruments that cover a wide frequency range to limit the mismatches in systematic errors introduced with combined data sets.

The work also provides an excellent demonstration that the capabilities of the SKA-Low precursor stations and pathfinder instruments (e.g. the MWA, LOFAR) can be employed to gain an improved understanding of pulsar radio spectra. A deeper analysis of the modelled spectra can be used to draw connections to the pulsar emission mechanism and to further constrain the location of the emission region. Moreover, the spectral modelling presented in this work can be expanded to a larger sample of pulsars to further our understanding of the low-frequency behaviour of pulsar spectra. These results will also inform pulsar population studies and make improved predictions of the expected yields of large pulsar surveys, which is particularly important in preparation for science with the SKA-Low (expected to be operational by the end of the decade).

\begin{acknowledgement}
We thank the referee for several useful comments that helped to improve this paper. We also thank Sam McSweeney for useful discussions. CPL was supported by an ICRAR Studentship and an EECMS Summer Scholarship through Curtin University. This scientific work makes use of the Murchison Radio-astronomy Observatory, operated by CSIRO. We acknowledge the Wajarri Yamatji people as the traditional owners of the Observatory site. We acknowledge the Pawsey Supercomputing Centre, which is supported by the Western Australian and Australian Governments. The AAVS2 and EDA2 are hosted by the MWA under an agreement via the MWA External Instruments Policy. The acquisition system was designed and purchased by INAF/Oxford University and the RX chain was design by INAF, as part of the SKA design and prototyping program. We acknowledge the support of the Curtin operations team, INAF group, and SKA-Low team in the development and on-going maintenance of the facilities used in this work. We acknowledge the use of the following software/packages for this work: {\sc psrchive} \citep{PSRCHIVE} and {\sc dspsr} \citep{DSPSR}; and the {\sc python} packages {\sc iminuit} \citep{iminuit}, {\sc psrqpy} \citep{psrqpy}, {\sc astropy} \citep{AstroPy}, {\sc matplotlib} \citep{Matplotlib}, and {\sc numpy} \citep{NumPy}. This work made use of NASA's Astrophysics Data System and arXiv.
\end{acknowledgement}

\bibliography{references}

\end{document}